\documentstyle[aps]{revtex}
\begin{document}
\draft
\preprint{CALT-68-1946}
\title{New Universality of the Baryon Isgur--Wise Form Factor
\\in the Large $N_c$ Limit}
\author{Chi-Keung Chow}
\address{California Institute of Technology, Pasadena, CA 91125}
\date{\today}
\maketitle
\begin{abstract}
New universality appears for the baryon Isgur--Wise form factor in the large
$N_c$ limit.
It is found that the semileptonic $\Lambda_b \rightarrow \Lambda_c$ and
$\Sigma_b^{(*)} \rightarrow \Sigma^{(*)}_c$ decays are described by the same
form factor, which can be calculated analytically.
In the exact chiral $SU(3)$ limit, the same form factor is applicable to
semileptonic $\Omega_b^{(*)} \rightarrow \Omega^{(*)}_c$ decays.
\end{abstract}

\narrowtext
In the heavy quark limit, the semileptonic $\Lambda_b \rightarrow \Lambda_c$
decay depends on a universal form factor $\eta(w)$ \cite{hb1,hb2,hb3,hb4,hb5},
which is defined by
\begin{equation}
\langle \Lambda_c (v',s')|\,\bar c \Gamma b\,|\Lambda_b (v,s)\rangle =
\eta(w)\, \bar u_{\Lambda_c}(v',s') \,\Gamma\, u_{\Lambda_b}(v,s) ,
\label{lf}
\end{equation}
where $w=v\cdot v'$.
The baryon Isgur--Wise form factor $\eta(w)$ is universal in the sense that it
depends on neither the masses nor the flavors of the heavy quarks.
Moreover, it does not depend on whether $\Gamma$ is a vector or an axial
vector.

Similar results hold for the semileptonic $\Sigma_b^{(*)} \rightarrow
\Sigma^{(*)}_c$ decay.
In this case we have two Isgur--Wise form factors, $\zeta_1(w)$ and
$\zeta_2(w)$ \cite{hb2,hb3}.
\begin{eqnarray}
&&\langle\Sigma^{(*)}_c (v',s')|\,\bar c \Gamma b\, |\Sigma^{(*)}_b(v,s)\rangle
\nonumber\\&&\; = (\zeta_1(w) g_{\mu\nu} + \zeta_2(w) v_\nu v'_\mu)\,
\bar u^\nu_{\Sigma^{(*)}_c}(v',s') \,\Gamma\, u^\mu_{\Sigma^{(*)}_b}(v,s) ,
\label{sf}
\end{eqnarray}
where $u^\nu_{\Sigma^*_b}(v',s')$ is the Rarita--Schwinger spinor vector for a
spin-$\textstyle {3\over2}$ particle.
$u^\mu_{\Sigma_b}(v,s)$ is defined by
\begin{equation}
u^\mu_{\Sigma_b}(v,s) = {(\gamma^\mu + v^\mu)\gamma_5 \over \sqrt 3 }
u_{\Sigma_b}(v,s)
\end{equation}
and similar for $u^\nu_{\Sigma^{(*)}_c}(v',s')$.

Recently, it has been realized that in the large $N_c$ limit, heavy baryons can
be viewed as the bound states of chiral solitons (N, $\Delta$) to heavy mesons
\cite{1,2,3}.
In this limit, a new universality of the baryon form factor appears; this
same form factor describes both the semileptonic $\Lambda_b \rightarrow
\Lambda_c$ and $\Sigma^{(*)}_b \rightarrow \Sigma^{(*)}_c$ decays, and both
$\zeta_1(w)$ and $\zeta_2(w)$ can be expressed in terms of $\eta(w)$.

We will describe the light degrees of freedom of a heavy baryon by $|I, a;
s_\ell, m\rangle$, where $I$ and $s_\ell$ are the isospin and spin of the light
degrees of freedom,  while $a$ and $m$ are their 3-components respectively.
Hence the light degrees of freedom of $\Lambda_Q$ is denoted by $|0, 0; 0, 0
\rangle$, while that of $\Sigma_Q$ is $|1, a; 1, m\rangle$.

Following the example of Ref. \cite{1}, the light degrees of freedom of a
chiral soliton is denoted by $|R, b; R, n)$.
A nucleon $N$ has $R = \textstyle {1\over2}$ while $\Delta$ has
$R = \textstyle {3\over2}$.
On the other hand, the light degrees of freedom of a heavy meson is denoted by
$|\textstyle {1\over2}, c; \textstyle {1\over2}, p\}$.

In the large $N_c$ limit, the binding potential between the chiral soliton and
the heavy meson is independent of both the isospins and the spins of the
particles \cite{3}.
We will denote the ground state wavefunction in momentum space as
$\phi(\bf q)$.
Then we have these decompositions:
\widetext
\begin{equation}
|0, 0; 0, 0\,(v)\rangle = \int d^3{\bf q} \,\phi({\bf q})\,
(\textstyle {1\over2}, b; \textstyle {1\over2}, c\,|0, 0)\;
(\textstyle {1\over2}, n; \textstyle {1\over2}, p\,|0, 0)\;
|\textstyle {1\over2}, b; \textstyle {1\over2}, n\,({\bf v}-{{\bf q}/ M_B}))\;
|\textstyle {1\over2}, c; \textstyle {1\over2}, p\,({\bf v}+{{\bf q}/ M_H})\} ,
\label{ld}
\end{equation}
\begin{eqnarray}
|1, a; 1, m\,(v)\rangle = \int d^3{\bf q} \,\phi({\bf q})\,
\Bigg[&&\sqrt{\textstyle{1\over3}}
(\textstyle {1\over2}, b; \textstyle {1\over2}, c\,|1, a)\;
(\textstyle {1\over2}, n; \textstyle {1\over2}, p\,|1, m)\;
|\textstyle {1\over2}, b; \textstyle {1\over2}, n\,({\bf v}-{{\bf q}/ M_B}))\;
|\textstyle {1\over2}, c; \textstyle {1\over2}, p\,({\bf v}+{{\bf q}/ M_H})\}
\nonumber\\+ &&\sqrt{\textstyle{2\over3}}
(\textstyle {3\over2}, b; \textstyle {1\over2}, c\,|1, a)\;
(\textstyle {3\over2}, n; \textstyle {1\over2}, p\,|1, m)\;
|\textstyle {3\over2}, b; \textstyle {3\over2}, n\,({\bf v}-{{\bf q}/ M_B}))\;
|\textstyle {1\over2}, c; \textstyle {1\over2}, p\,({\bf v}+{{\bf q}/ M_H})\}
\Bigg] .
\label{sd}
\end{eqnarray}
The $(j_1, m_1; j_2, m_2\,| J, M)$'s are the Clebsch--Gordon coefficients.
$M_B$ and $M_H$ are the masses of the chiral soliton and the heavy meson
respectively.

The spin-$\textstyle {1\over2}$ $\Lambda_Q$ is composed of a heavy quark with
spin-$\textstyle {1\over2}$ and light degrees of freedom with spin-0.
Hence
\begin{equation}
\langle \Lambda_c (v',s')|\,\bar c \Gamma b\,|\Lambda_b (v,s)\rangle =
\langle 0, 0; 0, 0\,(v')|0, 0; 0, 0\,(v)\rangle \bar u_c \,\Gamma\, u_b .
\end{equation}
Comparing with Eq. (\ref{lf}), we get \cite{3}
\begin{eqnarray}
\eta(w)&=&\langle 0, 0; 0, 0\,(v')|0, 0; 0, 0\,(v)\rangle
\nonumber\\
&=&\int d^3 {\bf q'}\,\int d^3 {\bf q}\,\phi^*({\bf q'})\,\phi({\bf q})\;
(\textstyle {1\over2}, b'; \textstyle {1\over2}, c'\,|0, 0)^*\;
(\textstyle {1\over2}, n'; \textstyle {1\over2}, p'\,|0, 0)^*\;
(\textstyle {1\over2}, b; \textstyle {1\over2}, c\,|0, 0)\;
(\textstyle {1\over2}, n; \textstyle {1\over2}, p\,|0, 0)
\nonumber\\& &
(\textstyle {1\over2}, b'; \textstyle {1\over2}, n'\,({\bf v'}-{{\bf q'}/ M_B})
|\textstyle {1\over2}, b; \textstyle {1\over2}, n\,({\bf v}-{{\bf q}/ M_B}))\;
\{\textstyle {1\over2}, c';\textstyle {1\over2}, p'\,({\bf v'}+{{\bf q'}/ M_H})
|\textstyle {1\over2}, c; \textstyle {1\over2}, p\,({\bf v}+{{\bf q}/ M_H})\} .
\label{lg}
\end{eqnarray}

Both the chiral soliton and the heavy meson matrix elements in Eq.(\ref{lg})
can be evaluated.
\begin{equation}
(\textstyle {1\over2}, b'; \textstyle {1\over2}, n'\,({\bf v'}-{{\bf q'}/ M_B})
|\textstyle {1\over2}, b; \textstyle {1\over2}, n\,({\bf v}-{{\bf q}/ M_B}))
= \delta_{bb'} \delta_{nn'} \delta^3({\bf v}-{\bf v'}-({\bf q}-{\bf q'})/ M_B),
\label{bm}
\end{equation}
\begin{equation}
\{\textstyle {1\over2}, c';\textstyle {1\over2}, p'\,({\bf v'}+{{\bf q'}/ M_H})
|\textstyle {1\over2}, c; \textstyle {1\over2}, p\,({\bf v}+{{\bf q}/ M_H})\}
= \delta_{cc'} \delta_{pp'} .
\label{hm}
\end{equation}
In Eq. (\ref{hm}) there should be an extra factor dependent on $\xi(w)$, the
meson Isgur--Wise form factor.
$\xi(w)$, however, is a slowly varying function and can be set to unity.

The Kornecker delta's made all Clebsch--Gordon coefficients vanish, and hence
\begin{equation}
\eta(w) = \int d^3 {\bf q}\, \phi^*({\bf q})\,
\phi({\bf q} + M_B({\bf v} - {\bf v'}))
\end{equation}
and the result in Ref. \cite{3} is reproduced.

Similarly, we can evaluate $\zeta_1(w)$ and $\zeta_2(w)$.
For concreteness we will focus on the decay $\Sigma^*_b\rightarrow\Sigma^*_c$.
The spin-$\textstyle {3\over2}$ $\Sigma_Q$ is composed of a heavy quark with
spin-$\textstyle {1\over2}$ and light degrees of freedom with spin-1.
Hence,
\begin{equation}
\langle\Sigma^{(*)}_c (v',s')|\,\bar c \Gamma b\, |\Sigma^{(*)}_b(v,s)\rangle
= \langle 1, a'; 1, m'\,(v')|1, a; 1, m\,(v)\rangle \bar u_c \,\Gamma\, u_b.
\end{equation}
The light matrix element can be evaluated in a way similar to Eq. (\ref{lg}).
\begin{eqnarray}
\langle 1, a';&& 1, m'\,(v')|1, a; 1, m\,(v)\rangle
= \int d^3 {\bf q'}\,\int d^3 {\bf q}\, \phi^*({\bf q'})\,\phi({\bf q})\;
\nonumber\\ \Bigg[ \textstyle {1\over3}&&
(\textstyle {1\over2}, b'; \textstyle {1\over2}, c'\,|1, a')^*\;
(\textstyle {1\over2}, n'; \textstyle {1\over2}, p'\,|1, m')^*\;
(\textstyle {1\over2}, b; \textstyle {1\over2}, c\,|1, a)\;
(\textstyle {1\over2}, n; \textstyle {1\over2}, p\,|1, m)
\nonumber\\&&
(\textstyle {1\over2}, b'; \textstyle {1\over2}, n'\,({\bf v'}-{{\bf q'}/ M_B})
|\textstyle {1\over2}, b; \textstyle {1\over2}, n\,({\bf v}-{{\bf q}/ M_B}))\;
\{\textstyle {1\over2}, c';\textstyle {1\over2}, p'\,({\bf v'}+{{\bf q'}/ M_H})
|\textstyle {1\over2}, c; \textstyle {1\over2}, p\,({\bf v}+{{\bf q}/ M_H})\}
\nonumber\\ \nonumber\\ + \textstyle {2\over3}&&
(\textstyle {3\over2}, b'; \textstyle {1\over2}, c'\,|1, a')^*\;
(\textstyle {3\over2}, n'; \textstyle {1\over2}, p'\,|1, m')^*\;
(\textstyle {3\over2}, b; \textstyle {1\over2}, c\,|1, a)\;
(\textstyle {3\over2}, n; \textstyle {1\over2}, p\,|1, m)
\nonumber\\&&
(\textstyle {3\over2}, b'; \textstyle {3\over2}, n'\,({\bf v'}-{{\bf q'}/ M_B})
|\textstyle {3\over2}, b; \textstyle {3\over2}, n\,({\bf v}-{{\bf q}/ M_B}))\;
\{\textstyle {1\over2}, c';\textstyle {1\over2}, p'\,({\bf v'}+{{\bf q'}/ M_H})
|\textstyle {1\over2}, c; \textstyle {1\over2}, p\,({\bf v}+{{\bf q}/ M_H})\}
\Bigg] .
\label{sg}
\end{eqnarray}
The cross terms vanish as
\begin{equation}
(\textstyle {3\over2}, b'; \textstyle {3\over2}, n'\,({\bf v'}-{{\bf q'}/ M_B})
|\textstyle {1\over2}, b; \textstyle {1\over2}, n\,({\bf v}-{{\bf q}/ M_B}))=0.
\end{equation}

As before, the chiral soliton and heavy meson matrix elements can be calculated
using Eq. (\ref{bm}), Eq. (\ref{hm}) and
\begin{equation}
(\textstyle {3\over2}, b'; \textstyle {3\over2}, n'\,({\bf v'}-{{\bf q'}/ M_B})
|\textstyle {3\over2}, b; \textstyle {3\over2}, n\,({\bf v}-{{\bf q}/ M_B}))
= \delta_{bb'} \delta_{nn'} \delta^3({\bf v}-{\bf v'}-({\bf q}-{\bf q'})/ M_B).
\end{equation}
By the identities
\begin{equation}
(j_1, m_1; j_2, m_2\,| J, M')^*\;(j_1, m_1; j_2, m_2\,| J, M) = \delta_{MM'} ,
\end{equation}
Eq. (\ref{sg}) can be simplified to
\begin{eqnarray}
\langle 1, a'; 1, m'\,(v')|1, a; 1, m\,(v)\rangle &=& \delta_{aa'} \delta_{mm'}
\int d^3 {\bf q}\, \phi^*({\bf q})\,\phi({\bf q} + M_B({\bf v} - {\bf v'}))
\nonumber\\ &=& \delta_{aa'} \delta_{mm'} \eta(w) .
\end{eqnarray}
Notice that $\eta(w)$ has reappeared.
The expected $\delta_{aa'}$ is just a consequence of isospin conservation in
the weak decay.
On the other hand, the $\delta_{mm'}$ demands that the initial and final light
degrees of freedom are in the same spin state.
In term of the polarization vectors $\epsilon$ and $\epsilon'$ we have
\begin{equation}
\langle\Sigma^*_c (v',\epsilon',s')|\,\bar c \Gamma b\,|
\Sigma^*_b(v,\epsilon,s)\rangle={\eta(w)\over1+w}\bigg[(1+w)g_{\mu\nu}
-v_\nu v'_\mu\bigg]\epsilon'^{*\nu} \epsilon^\mu \bar u_c\,\Gamma\,u_b.
\label{s}
\end{equation}
The combination in the square brackets should be familiar as it also appears
in the $B \rightarrow D^*$ decay.
\begin{equation}
\langle D^*(v',\epsilon')|\,\bar c \gamma_\mu\gamma_5 b\,|B(v)\rangle
= \xi(w) \bigg[(1+w)g_{\mu\nu}-v_\nu v'_\mu\bigg]\epsilon'^{*\nu} .
\end{equation}

\narrowtext
Comparing Eq. (\ref{s}) to Eq. (\ref{sf}), we conclude that, in the large $N_c$
limit,
\begin{equation}
\zeta_1(w) = -(1+w)\zeta_2 = \eta(w) .
\label{pdn}
\end{equation}

Thus the main result of this paper is obtained.
The same universal form factor $\eta(w)$ describes the weak decays of
$\Lambda_Q$ and $\Sigma_Q$.
In the real world, $\Sigma^{(*)}_b$ decays strongly and the decays
$\Sigma^{(*)}_b \rightarrow \Sigma^{(*)}_c$ are hardly ever observed.
On the other hand, $\Omega_b$, the chiral $SU(3)$ partner of $\Sigma_b$, does
decay weakly to $\Omega^{(*)}_c$.
Then chiral $SU(3)$ predicts that the $\Omega_b \rightarrow \Omega^{(*)}_c$
decay is also described by the same universal form factor $\eta(w)$.
This statement may be put to experimental test in the future.
The deviation from Eq. (\ref{pdn}) is an indication of how good the large $N_c$
limit is.

The exact form of $\eta(w)$ depends solely on $\phi({\bf q})$.
In the large $N_c$ limit, the binding potential between the heavy meson and
the chiral soliton is simple harmonic \cite{3}.
\begin{equation}
\phi({\bf q})={1\over(\pi^2 M_B\kappa)^{3/8}}\,
\exp\left(-{\bf q}^2\big/2\sqrt{M_B\kappa}\right)
\end{equation}
in the heavy quark limit.
The parameter $\kappa$ is the spring constant of the simple harmonic potential,
which is independent of the heavy quark species (by heavy quark symmetry) and
the baryon spin (by the large $N_c$ limit).
Its value can be determined to be $(530 MeV)^3$ in the Skyrme model and
$(440 MeV)^3$ from $\Lambda^*_c-\Lambda_c$ splitting.

With this particular form of $\phi({\bf q})$, $\eta(w)$ can be calculated.
\begin{equation}
\eta(w) = \exp\left(-(w-1)\sqrt{M_B^3\big/\kappa}\right) .
\end{equation}

Similarly, we can generalize our treatment to weak form factors for decays to
orbitally excited baryons.
It has been shown in a recent paper \cite{lp} that both the Bjorken
\cite{lb1,lb2} and Voloshin \cite{lv} sum rules for $\Lambda_b$ decays are
saturated by the first doublet of excited $\Lambda_c$ states.
In fact heavy quark symmetry decrees that a single form factor describes all
$\Lambda_b$ decays to this doublet, and this form factor is (up to some
kinematic factors) just
\begin{equation}
\eta_1(w)=\int d^3 {\bf q}\, \phi_1^*({\bf q})\,
\phi({\bf q} + M_B({\bf v} - {\bf v'})) ,
\end{equation}
where $\phi_1({\bf q})$ is the momentum wavefunction of the first orbital
excited state of the simple harmonic potential.

It is straightforward to repeat the exercise for $\Omega_b$ decays.
Heavy quark symmetry allows four independent form factors for $\Omega_b$ decays
into the first excited $\Omega_c$ pentalet \cite{sb}.
Yet in the large $N_c$ limit, all of them are expressible in $\eta_1(w)$.
Likewise, both the Bjorken \cite{sb} and the Voloshin sum rules are again
saturated by this pentalet.

It should be noted that Eq. (\ref{s}) does not depend on any particular form of
$\phi({\bf q})$ or $\eta(w)$.
It does depend, however, on the crucial assumption that the binding potential
is independent of the isospins and spins of the particles.

We have proved that, in the large $N_c$ limit, a new universality appeared for
the baryon weak form factor $\eta(w)$.
This same form factor describes weak decays of both $\Lambda_Q$ and $\Omega_Q$
in the exact chiral $SU(3)$ limit.
In the real world, where chiral $SU(3)$ is broken, the leading correction is
expected to come from kaon loops.
In particular, it is expected that, near the point of zero recoil
\begin{equation}
|\zeta_1(w)-\eta(w)|\sim{g^2\Delta^2\over{(4\pi f)}^2}
\ln\left({m_K^2\over\mu^2}\right) ,
\end{equation}
where $g$ is the pion-heavy meson coupling constant and $f$ the pion decay
constant.
$\Delta$ is the $\Omega^*_Q-\Omega_Q$ splitting and $\mu$ the subtraction
point.
The correction is expected to be about 1\%, but the relevant loop integrals
must be calculated and the counterterms known to get the exact magnitude of the
correction.

This work was supported in part by the U.S. Dept. of Energy under Grant No.
DE-FG03-92-ER 40701.

\end{document}